\documentclass[preprint,1p,authoryear]{elsarticle}

\usepackage{amssymb}
\usepackage{amsmath}
\usepackage[colorlinks=true, linkcolor=blue, citecolor=blue, urlcolor=blue]{hyperref}
\usepackage{balance}
\usepackage{graphicx}
\usepackage{booktabs}
\usepackage{multirow}
\usepackage{threeparttable}
\usepackage{url}

\begin{document}

\title{Crisis Communication in the Face of Data Breaches}
\author[sdu]{Jukka Ruohonen\corref{cor}}
\ead{juk@mmmi.sdu.dk}
\author[utu]{Kalle Hjerppe}
\author[jyu]{Katleena Kortesuo}
\cortext[cor]{Corresponding author.}

\address[sdu]{University of Southern Denmark, DK-6400 S\o{}nderborg, Denmark}
\address[utu]{University of Turku, 20014 Turun yliopisto, Finland}
\address[jyu]{University of Jyv\"askyl\"a, PL 35, 40014 Jyv\"askyl\"an yliopisto, Finland}

\begin{abstract}
Data breaches refer to unauthorized accesses to data. Typically but not always,
data breaches are about cyber crime. An organization facing such a crime is
often also in a crisis situation. Therefore, organizations should prepare also
for data breaches in their crisis management procedures. These procedures should
include also crisis communication plans. To this end, this paper examines data
breach crisis communication strategies and their practical executions. The
background comes from the vibrant crisis communication research
domain. According to a few qualitative case studies from Finland, the
conventional wisdom holds well; the successful cases indicate communicating
early, taking responsibility, offering an apology, and notifying public
authorities. The unsuccessful cases show varying degrees of the reverse,
including shifting of blame, positioning of an organization as a victim, and
failing to notify public authorities. With these qualitative insights, the paper
contributes to the research domain by focusing specifically on data breach
crises, their peculiarities, and their management, including with respect to
European regulations that have been neglected in existing crisis communication
research.
\end{abstract}


\maketitle

\section{Introduction}

A data breach can be defined simply as unauthorized access to data. The data
involved may be personal data of natural persons, but it may also refer to
valuable business data or sensitive and even classified data held by public
sector organizations. Therefore, data breaches pose a significant reputation
risk to organizations. Financial consequences are likely for companies. If a
data breach has involved personal data, also these victims may be exposed to
financial fraud in addition to psychological harm and even threats to personal
safety. Data breaches also reduce trust throughout a society.

An organization may face different crises throughout its lifetime. What
separates an organizational crisis from other unpleasant situations is the
threat to an organization's valuable assets, the surprise element, meaning that
the timing of a crisis is almost always unexpected or unanticipated, and the
urgency and short response time~\citep{Bell10, Ndlela19}. Organizational crises
are low-probability but high-impact events~\citep{Syed19}. Then, crisis
management refers to the processes by which organizations deal with crises
before, during, and after their occurrence~\citep{Badu23}. These
characterizations apply also to data breach crises. Nowadays, data is a valuable
asset for most organizations, and breaches of data thus threaten organizations'
critical assets. Even under careful cyber security preparations, data breaches,
like most successful cyber attacks, are also unexpected and extraordinary events
to an organization. Like all organizational crises, data breaches further
require urgent attention from an organization.

Before a crisis, to prevent data breaches from occurring to begin with,
organizations should invest in cyber security. It is impossible to provide
universal guidelines on which areas the investments should specifically focus,
but typical examples include sound engineering practices, testing and hardening,
patching procedures for known vulnerabilities, documentation including an
incident response plan, auxiliary technical solutions such as intrusion
detection systems, training for employees and security awareness campaigns, and
even so-called bug bounties and red team exercises. However, not all data
breaches occur due to technical exploitation of vulnerabilities in systems; also
so-called inside jobs and lost or stolen computer equipment have been relatively
frequent causes for data breaches~\citep{Lanois19, Ruohonen24IWCC}. Therefore,
the investments should target also organizational security. In any case, an
organization faces many tasks also during and after a data breach crisis. While
it is again impossible to list all potential tasks, the typical examples include
damage examination and minimization of further damage, setting up a team of
expert investigators, informing those affected, in-house legal counseling, and
contacting of an insurance company, among other things~\citep{Teichmann24}. The
informing obligation falls to the domain of crisis communication, which too is a
central element in crisis management. Ideally, an organization should also have
a crisis communication strategy before a data breach crisis occurs.

In fact, crisis communication is not an optional procedure; it is a
necessity. If an organization's goal is to comply with the information security
standard ISO/IEC 27000, it must also plan crisis communication. The standard
requires, among other things, that information security instructions are
available to all employees and that possible deviations can be reported quickly
within an organization \citep{Diamantopoulou20, Kamil23}. In practice, this
requirement is preventive crisis communication; clear information security
instructions and open reporting of deviations help to prevent crises and reduce
their impact. In addition, the standard requires a crisis communication plan for
data breaches and data leaks. A complying organization must also have ready-made
communication models, roles, and structures because in case of a disruption it
must be able to communicate with public authorities, customers, stakeholders, and
media~\citep{Diamantopoulou20}. Regulations add further requirements.

While there are some existing studies on crisis communication in data breach
crises situations, the amount of this literature is very modest when compared to
the more general crisis communication research literature. Therefore, more
contributions to this underresearched domain are always welcome. The literature
contains some notable limitations. Among these is a lack of research that would
have connected data breach crisis communication to existing regulations. To this
end, this paper discusses data breach crises and their communication in a
European context. The two European regulations considered are the General Data
Protection Regulation (GDPR) and the so-called NIS2 directive, that is,
Regulation (EU) 2016/679 and Directive (EU) 2022/2555, respectively. The former
became enforceable in 2018, while the latter should be transposed by the member
states by October 2024. While both laws impose requirements for incident
response, among other things, particularly the GDPR is important in the present
context because it contains many legal requirements for handling personal data
breaches, including legal liability from careless or negligent information
security practices.

These regulations and their relation to crisis communication are discussed in
the opening Section~\ref{sec: crisis communication}. The section also serves as
a theoretical overview and a literature review. The overall goal is to better
connect the limited and specific data breach literature to the domains of
strategic communication and public relations. A further notable limitation in
the existing crisis communication research literature is that it has mostly
concentrated on private sector companies~\citep{Horsley02}. However, data
breaches can occur to any type of an organization, including public sector and
non-governmental organizations. To address this limitation, after a brief
elaboration of materials and methods in Section~\ref{sec: materials and
  methods}, the subsequent Section~\ref{sec: case studies} presents eight short
qualitative case studies that contain also public sector data breaches. The goal
is to examine the crisis communication strategies employed by the organizations
affected by data breaches; these strategies range from apologies to shifting of
blame, offering compensations, and communicating early. Finally,
Section~\ref{sec: discussion} provides a concluding discussion, an elaboration
of limitations, and some ideas for further research.

\section{Crisis Communication and Data Breaches}\label{sec: crisis communication}

\subsection{Images, Situations, and Strategies}

The two presumably most influential theories in crisis communication are the
image repair theory (IRT) and the situational crisis communication
theory (SCCT). The first is based on the assumption that crisis communication is
primarily about reducing reputation damage. This goal-oriented theory comes with
different communication strategies for repairing an image, including defensive
utterances, such as justifications, denials, excuses, and apologies, and other
rhetorical and persuasive tactics~\citep{Benoit15}. By the means of
communication and impression management, an organization affected by a crisis
should thus seek to quickly reestablish the image, reputation, and legitimacy of
the organization~\citep{Bell10}. Such reestablishment is often easier when an
organization already has a good reputation~\citep{Nohammer23}. Therefore,
impression management should apply throughout an organization's lifetime. Once a
crisis is over, it is possible to foster the image restoration process by
renewal discourse, emphasizing the challenges, opportunities, and lessons
learned~\citep{Liu20}. The situational theory variant is based on the same
rationale of image preservation or restoration. Also it comes with different
crisis communication strategies, but with an addition of guidelines on how to
match the strategies to different types of crises and their situational
characteristics, including how much an organization is responsible, the
organization's ability to handle a given crisis, and the severity of the
crisis~\citep{Coombs02}. The situational characteristics are important because
these underline that an organization may be guilty of wrongdoing, but
organizations often face also different crises that are beyond their control,
whether these are natural disasters, destructive fires, major labor union
strikes, terror attacks, or other \textit{force majeure} events. Obviously,
therefore, also the communication strategies should differ.

According to existing results, having a strategy is beneficial to an
organization; it helps to maintain customer support during and after a
crisis~\citep{Racer01}. However, the evidence is generally mixed regarding the
universal appropriateness of some particular crisis communication strategies. If
an organization is responsible for a crisis, some authors maintain that a decent
strategy involves apologizing, accepting responsibility, and showing
remorse~\citep{Claeys16, Horsley02}. This strategy can be accompanied with other
related strategies, such as bolstering of positive information to the crisis
situation and redress even in the form of monetary
compensations~\citep{Coombs15}. Denial and attempts to shift blame to others can
quickly backfire like many of the other more offensive strategies, such as
deceptive spins, downplaying, or attacking the accuser.  In fact, backfiring is
guaranteed in case an organization is later found to bear responsibility for a
given crisis. However, denial can still work in low-risk and low-responsibility
situations, including the cases in which an organization already has a bad
reputation, and apologies are culturally dependent; not all apologies are the
same and these may be perceived as insincere in some settings, and there may
even be legal repercussions from outright accepting liability~\citep{Coombs02,
  Park23, Nohammer23}. Denial is also an appropriate strategy in so-called
non-crisis situations; sometimes it may be that an incident is merely a rumor
spread by some malicious parties and people on social media. Although litigation
is slow and faces many challenges in the data breach context~\citep{Mills17},
for better or worse, fear of litigation has still been a typical reason to limit
data breach crisis communication~\citep{Thomas22}. These considerations justify
the SCCT; there is no single strategy that would work in all cases.

Both theories are relevant also for data breaches, which usually involve a high
risk of reputation damage to organizations. Data is the new oil, so the saying
goes, and thus losing valuable data to criminals is hardly good business to say
the least. Data breaches also demonstrate how a company's reputation may have
direct financial consequences. A~data breach may well decrease consumers'
purchasing intentions. A~grave data breach may even affect the stock price of a
company~\citep{Johnson17}, and a large espionage scandal may affect a whole
country's image, and so forth.

Regarding communication strategies, it seems sensible to recommend that
organizations should stick to facts and truth-telling as closely as
possible. For reasons soon to be discussed, facts are likely to emerge sooner or
later in any case. However, this strategy does not imply that all facts should
be disclosed to the public. For instance, sensitive technical details should be
left out already because these might expose an organization to further attacks
and incidents~\citep{Thomas22}. Provided that a data breach is real and not
merely a rumor, strategies such as denial and justification are likely bad
choices at least in the European context. While denial is problematic because of
the backfiring risk, justification is likely a poor choice because it will
explicitly signal the poor cyber security posture of an organization. These
points do not mean that the conventional wisdom should be abandoned
altogether. Indeed, apologies and admission of responsibility are expected after
a data breach, and compensations may further reduce the reputation
damage~\citep{Jenkins14, Knight20, Kuipers22, Masuch21}. Regarding the facts to
be disclosed, at least the number of people affected should be
delivered~\citep{Chatterjee19}. It is further important to disclose the type of
personal data that has been breached. Such disclosure is often necessary for the
people affected to protect themselves. Without knowing whether social security
numbers have been breached, for instance, it is impossible to prepare for a
threat of identity thefts. An analogous point applies to financial
identifiers such as credit card~details.

In contrast, softer strategies such as storytelling~\citep{Lee20a} and other
narrative approaches may have a limited applicability in the data breach
context. Moreover, the more there are facts, the less room there is for
narratives and stories~\citep{Clementson20b}. Rather analogous points apply with
respect to dealing the public's emotional response. Data breaches are oftentimes
emotionally intensive for the people affected; these often arouse strong
negative feelings, such as anxiety, fear, disgust, sadness, and
anger~\citep{Bachura17, Chatterjee19, Syed19}. Strong negative emotions are also
likely to increase the probability of negative word-of-mouth after a
crisis~\citep{Coombs15}. Existing results further indicate that particularly
spinning stories anger people further, while giving answers without
embellishment lessens the public's anger toward an
organization~\citep{Clementson20a}. Such results bespeak for the plain
truth-telling as a strategy.

If narrative approaches are still used, a possible strategy involves trying to
frame the organization affected by a data breach as a victim too, possibly
further offering mitigative instructions for the people affected in order to
lessen their anger, disgust, and other negative emotions~\citep{Syed19}. The
mitigative instructions may involve security training material, but an
organization affected by a data breach might offer also free protective
services, such as credit monitoring or fraud protection
solutions~\citep{Borgesius23, Chatterjee19}. While some authors have been
critical about this strategy, arguing against ``playing the victim
card''~\citep{Knight20, Kuipers22}, the strategy is sincere in the sense that
many organizations affected by data breaches truly are victims of computer
crimes. At the same time, many of these victim organizations have also suffered
from poor defensive cyber security or even plain negligence thereto, which may
well decrease the strategy's effectiveness compared to the conventional strategy
of admitting responsibility, offering apology, and providing redress or
mitigations. Because the strategy typically leads to two victim types, the
organization and the persons affected, it may also complicate or even prevent
other mitigative strategies, such as offering sympathy to the persons affected,
which has been observed to lessen negative emotions~\citep{Xu20}. A similar
maneuver does not work well with organizations because it is difficult to feel
sympathy or empathy toward an abstract entity such as an
organization~\citep{Schoofs22}. The duality may cause also other problems. For
instance: if an organization positions itself alone as a victim, the other
victims, the persons affected may show progressive disappointment and anger
toward the organization. These points again underline that while the literature
offers some practical recommendations and best practices, there is no single way
to communicate in a data breach crisis.

\subsection{Informing People and Stakeholders}

The IRT and SCCT may sound a little cynical, as both are primarily about
protecting an organization and its image with communication strategies, some of
which may even be seen as borderline unethical. Therefore, it should be
emphasized that crisis communication involves also informing the victims of a
crisis. In fact, it can be argued that the foremost task in reliable crisis
communication is to inform the public about an incident and the ways people
affected may protect themselves~\citep{Lee20a}. The public should know what is
happening, who is affected, and what countermeasures the people affected may
take. This requirement is particularly important in serious, life-threatening
crises, such as natural or man-made environmental catastrophes and public health
emergencies. That said, the requirement applies also to data breach crises. As
already said, immediately informing the people affected is important in order
for them to prepare themselves for countering identity thefts and other harms;
these victims do not have the time to wait for the legal system and regulators
to give them adequate relief~\citep{Mills17}. To this end, some have recommended
a dialogue approach with the public, arguing against a prevailing myth that the
public will panic or behave inappropriately due to the information
delivered~\citep{Chatterjee19, Seeger06}. Rather, panic tends to occur when a
crisis is suddenly communicated by third-parties such as the
media~\citep{Badu23}. To be effective, such a dialogue approach requires that an
organization already knows where the public stands; therefore, media monitoring,
public opinion surveys, and related techniques should be preferably applied
prior to a crisis.

Similar points apply with respect to informing an organization's important
stakeholders, whether they are employees, financiers, suppliers, interest
groups, or governmental actors. In fact, it can be argued that stakeholders
should be informed prior to informing the public, as stakeholders are vital to
an organization's operating environment. A~failure to inform stakeholders may
cause long-lasting trust issues to an organization.

Also the communication strategies may differ between the public and
stakeholders. If there is confidence in the stakeholders, such that no
information leaks can be assumed to occur, it is often recommended to stick as
close to the truth as possible, providing information that is reliable,
consistent, and credible~\citep{Ndlela19}. The information delivered may include
also soft guidelines on how a stakeholder should behave and what is expected
from the stakeholder in the crisis situation~\citep{Marynissen20}. Because
coordination can be defined as a process of managing dependencies between
activities~\citep{MaloneCrowston94, Ruohonen18IST}, such guidelines may reduce a
risk that coordination fails because stakeholders do not understand their
actions and the impact of these actions to other stakeholders and the crisis
itself~\citep{Badu23}. Regarding communication, it can indeed be detrimental to
an organization's image if stakeholders communicate on their own with diverging
messages. Particularly public disagreements should be
avoided~\citep{Knight20}. The more there is convergence in the messages, the
more neutral and credible the sender~\citep{LuomaAho17}. Actually, through
inter-organizational cooperation, it is possible to strengthen a message
delivered to the public by combining the voices from multiple
stakeholders~\citep{Liu20}. Again, an obvious prerequisite is that an
organization knows all its relevant stakeholders; techniques such as stakeholder
mapping can be used for this task. Also trust-building should apply among
stakeholders prior to a crisis because cooperation and coordination are known to
work poorly in the absence of trust~\citep{Badu23}. The importance of
stakeholders is further reinforced by regulations.

If a data breach has involved personal data, like most data breaches likely
have, the organizations who process personal data on behalf of an organization
who possesses the data should notify the controller according to the GDPR's
Article~33. In other words, a data breach affecting a processing stakeholder
should be communicated to the parent organization who holds the personal
data. This mandate is important because it should in theory ensure that
important information traverses in supply-chain cyber attacks involving personal
data. Such attacks have recently gained prominence~\citep{Martinez21,
  YeboahOfori19}. Although the GDPR does not consider the reverse relation, also
it should be taken into account already due to maintaining trust between
business associates or other important stakeholders who deal with personal
data. If there are multiple possessors, the responsibilities and compliance
should be negotiated in advance according to Article 26 in the regulation. These
regulatory mandates again signify the importance of prior planning. Ideally,
this planning should include also joint crisis communication plans among the
various possible processors and possessors. Without plans, it is possible that
so-called crisis spillover will occur, meaning that also a victim organization's
stakeholders and partners will suffer from reputation damage or other negative
consequences~\citep{Voges24}. Furthermore, the information exchanges are not
limited to traditional stakeholders and business partners. Also regulators and
governmental bodies should be typically notified after a data breach.

\subsection{Notifying Public Authorities}\label{subsec: notifying public authorities}

Today, in Europe, regulators and governmental actors are particularly
relevant. If a data breach has involved personal data, a national data
protection authority (DPA) should be informed. According to Article~33 in the
GDPR, omission of the notification is possible only in case a breach ``is
unlikely to result in a risk to the rights and freedoms of natural
persons''. Personal data, as defined in the GDPR's Article~4, refers to ``any
information relating to an identified or identifiable natural person''. The
scope is thus much wider than the personally identifiable information (PII)
concept used in the United States. While a risk assessment is left to an
organization affected by a personal data breach, most breaches therefore should
involve a notification to a DPA~\citep{Kasl21}. Without undue delay,
according to Article~34 in the GDPR, a high-risk personal data breach should be
further communicated to the persons affected either directly or through mass
communication. According to the same article, a DPA may also force an
organization to comply with this informing obligation toward the people affected
by a personal data breach. To determine a high-risk scenario, an organization
might deduce whether a breach has involved special categories of personal
data~\citep{Borgesius23}, as defined in the GDPR's Article 9. Though, arguably
also leakage of conventional personal identifiers, such as social security
numbers, satisfy a high-risk scenario for the persons affected, as also
  hinted by the GDPR's recital 75. In addition, the GDPR imposes many other
requirements, including documentation about personal data processing, its legal
basis, and technical and organizational security measures to protect the data
and its processing. The documentation should be in place even in case an
organization determines that a breach does not pose a high risk to the natural
persons affected. According to Article 33, detailed documentation should be
available to a DPA also with respect to already occurred personal data breaches.

Furthermore, the NIS2 directive has laid down notification obligations for
so-called essential and important entities. These entities include not only
traditional critical infrastructure providers, such as those operating in the
energy and transport sectors, but also the banking, finance, healthcare, water,
and food sectors are covered together with public administration units and some
of the Internet's core infrastructure elements. The notifications required
should be delivered to national computer security incident response teams
(CSIRTs), which are distinct from DPAs in most member states. However, according
to Article 23 in the NIS2 directive, only significant incidents should be
reported; these include those incidents that cause severe operational disruption
or those that have affected or are capable of affecting ``other natural or legal
persons by causing considerable material or non-material damage''. Thus, also
severe data breaches are covered, including large-scale personal data breaches
and serious breaches involving non-personal data. If a data breach has involved
a computer crime, such as an intrusion to systems or networks, like most data
breaches have, a criminal report should be further made to a national law
enforcement unit.

After having potentially contacted one or more of these three public authority
types, the communication is not any more entirely in the hands of an
organization itself. While confidentiality should be guaranteed with the public
authorities, it is clear that some of the crisis communication strategies, such
as outright denial, are no longer applicable after this point. Thereafter, the
diversity of voices in media may increase as public authorities may also appear
in media, providing expertise and their own interpretations of a given data
breach crisis, which often focus (or should focus) on those affected because
public authorities first and foremost represent the public
interests~\citep{Raupp19}. To decrease the diversity, which may have negative
consequences for an organization's image, it may be possible in some serious
cases to continue crisis communication jointly with the public authorities
involved. Another related option is to align an organization's data breach
crisis communication to the voice from public authorities.

It has been suspected that some of the somewhat shady communication tactics,
such as controverting and delaying of communication, have decreased after the
GDPR was enacted~\citep{Shastri21}. As a data breach crisis unfolds, it is
indeed often recommended to remain open and transparent with both regulators and
the public~\citep{Lanois19}. Both the GDPR and the NIS2 directive entail also
potentially large administrative fines from non-compliance or other misbehavior,
which should further increase the incentive for remaining transparent and
trustworthy. Indeed, cyber security incidents, including data breaches, have
been a frequent reason for GDPR enforcement fines, some of which have been
levied also due to a failure to comply with the GDPR's notification
mandates~\citep{Ruohonen22IS}. If a breach leads to a formal investigation by a
DPA or other public authority, also the crisis communication strategy should
likely further change. Under these settings, a strategy of acknowledging and
waiting may work; an organization should acknowledge a data breach and
communicate it to the persons affected, as also required by the GDPR in
high-risk personal data breaches, but then wait for the results of an
investigation before communicating further~\citep{Jong20}. To some extent, this
strategy aligns with a ``no comment'' strategy often advised by legal
counsel~\citep{Horsley02}. These strategies may even be a necessity in some
cases. For instance, a law enforcement unit may forbid a disclosure of any
sensitive details during a criminal investigation of a data breach.

\subsection{Timing of Communication}

Timing of crisis communication is important. During the start of a crisis, there
is usually a very high demand for information from the press and the people
affected by the crisis. Therefore, it is usually maintained that delaying
communication is a bad idea; waiting to address audiences and the media at a
later time when more information is available can be detrimental to an
organization's image~\citep{Horsley02}. The reason is clear: a lack of
communication often leads to rumors and speculations about the organization in
question, perhaps suggesting that it may have something more to hide or that it
simply does not care. Such rumors and speculations nowadays spread fast and wide
due to social media. Thus, traditional framing and perception shaping strategies
may work poorly in the modern digital era in which multiple online media and
so-called social media influencers may produce their own crisis
narratives. Influencers may also significantly contribute to the public's
emotional response~\citep{Mak19}. Therefore, analogously to the earlier points
about stakeholders and public authorities, an organization should communicate
early in order to root a desired crisis narrative. As said, early communication
is also necessary for the persons affected to protect themselves.

Rapid response is behind the common idiom of ``stealing the thunder'', meaning
generally that organizations should self-disclose incriminating information
instead of trying to conceal it. Such self-disclosure may affect positively the
image of an organization facing a crisis because the organization may be
perceived as being honest, credible, and capable of handling the
crisis~\citep{Lee20b}. In reverse, inactivity and passivity tend to be
associated with ineffectiveness and a lack of trust~\citep{LuomaAho17}. The
actual response strategy involves two steps: the self-disclosure or ``stealing''
should occur before the media and public discover a crisis, and, then,
``thunder'' occurs when an organization waits for and responds to inquiries from
the media, stakeholders, and public~\citep{Claeys16}. It is possible to
incorporate other strategies, such as apologies, to the response phase. This
strategy presumably works better when an organization has already built a
relationship with media beforehand. Regardless, once the thunder has arrived, it
is important for an organization's spokespersons to remain accessible to media;
there should be no information vacuums~\citep{Ndlela19}. Then, during
communication with media, it is often advised to avoid inconsistency by
accepting uncertainty and avoiding offers of overly reassuring
messages~\citep{Seeger06}. This strategy may work also in the data breach
context.

In particular, as could be expected, proactive self-disclosure has helped
organizations affected by data breaches to control the narratives in media,
whereas communication delays have often caused harsh criticism in
media~\citep{Kuipers22}. Of course, stealing the thunder strategy requires that
an organization is the first entity to become aware of a data breach and to
communicate it to the media and public. This prerequisite can be challenging in
the data breach context because data breaches are difficult to detect within an
organization; typically, it can take months before a data breach is
detected~\citep{Ruohonen24IWCC}. Therefore, at least the criminals involved
already possess the data stolen and the associated critical information. If the
criminals have not made the data and information public, the incentive for
stealing the thunder should increase. Because a relatively large amount of data
breaches is detected by third-parties instead of the organizations
affected~\citep{Weir17}, the incentive should intensify since information about
a data breach may already be located outside of an organization. Even when the
third-parties involved are not malicious, a leakage of the sensitive information
about the existence of a data breach may occur.

Also regulations impose strict requirements for the timing of
communication. According to the GDPR's Article~33, a data breach involving
personal data should be communicated to a DPA no later than 72 hours after
having became aware of it. Furthermore, according to Article~23 in the NIS2
directive, essential and important entities should forward an early warning
about a significant incident to a CSIRT within 24 hours of becoming aware of
such an incident, and this notification should be accompanied by a more
comprehensive incident notification within 72 hours. Thereafter, a CSIRT may
request status updates at any time. Furthermore, within a month, a final report
of the significant incident should be delivered. Thus, in Europe the urgency in
data breach crises situations is intensified by regulations. The very strict
deadlines and short response times, which have been also
criticized~\citep{Borgesius23, Kasl21, Weir17}, yet again signify the
importance of having crisis communication and incident response plans in
advance. Although the GDPR does not specify a strict deadline for informing the
people affected by a high-risk personal data breach, it seems sensible to
recommend that stealing the thunder, if not already done, should occur fairly
quickly after a notification has been delivered to a DPA. The same point applies
with respect to the significant incidents specified in the NIS2 directive. All
in all, also timing should belong to a data breach crisis communication plan.

\section{Materials and Methods}\label{sec: materials and methods}

The concise qualitative analysis operates with eight data breach cases. As a
background, it should be recalled that case selection in qualitative research
differs from statistical sampling approaches typically used in quantitative
research. Three criteria were used to select the cases. In addition to (a) the
public sector criterion noted in the introduction, the case selection was done
so that (b) both good and bad, or successful and unsuccessful,
\textit{illuminative} cases are present~\citep{Meredith98, Suri11}. Although
generalizability remains an issue, as is usual in qualitative research, this
so-called extreme or deviant case sampling based on polar types is useful and
appropriate to qualitatively analyze the contextual factors that shaped the
successes and failures~\citep{Eisenhardt89}. In addition, (c)~all cases are from
Finland. This geographic framing is generally welcome because most of the
existing research on data breaches has concentrated on large breaches
particularly in the United States.

The empirical material for all cases is based on news articles in mainstream
media, press releases from the organizations affected, and official statements
from public authorities, including law enforcement and the Finnish DPA in
particular. While hundreds of news articles, press releases, and statements were
examined and collated, in what follows, explicit references to these are omitted
for brevity. All materials examined and used were also in Finnish, and thus
references to these have limited appeal for an international audience. More
importantly, therefore, it is relevant to emphasize that the empirical material
reflects how the organizations' external environments responded to their crisis
management, including their crisis communication. This viewpoint is important
because media and journalism play an important role in data
breaches~\citep{Ruohonen24IWCC}. It could be even argued that crisis
communication in this context is largely a continuous and iterative tango with
media; particularly missteps and mistakes in crisis communication quickly find
responses in media, which, in turn, feed into the responses and reactions of the
general public.

The continuous nature of crisis communication in the data breach context implies
that the associated crisis management should be examined as a process. A
procedural approach is necessary also for knowing whether the regulations
discussed in the previous section were followed, and what the potential
responses from public authorities and other stakeholders were. To these ends,
qualitative process tracing is used as a methodology. While rich both in history
and variations, qualitative process tracing generally seeks to retrospectively
identify potential causes that produced given outcomes~\citep{Beach20}. It
aligns well also with the case selection based on polar types. As the outcomes
are known, even though only based on subjective evaluation, a particular
interest is whether the theoretical premises in crisis communication and the
practical advices from the literature were followed by the organizations
examined in the case studies. In addition, the tracing can reveal whether these
premises and advices have missed important elements that increase the likelihood
of a success or a failure. It is also possible to evaluate whether some
theoretical premises really are necessary conditions for a success. In process
tracing this evaluation amounts to examining whether given causal factors
hypothesized actually occurred~\citep{Mahoney12}. Although the theoretical
premises and practical advices are used as premises for deductive reasoning, the
tracing approach is still more on the side of inductive theory-building than on
the side of strict confirmatory analysis. Given the small sample size and the
qualitative approach taken, the goal is thus to provide narratives that
differentiate the major sequences of the overall processes, identifying critical
moments and factors that shaped the processes and their
outcomes~\citep{Blatter14}. The presentation of the case study results reflects
this goal; the breaches are elaborated as concisely as possible, focusing on the
critical moments and factors.

\section{Case Studies}\label{sec: case studies}

\subsection{The Vastaamo Breach}\label{subsec: vastaamo}

The Vastaamo data breach shocked Finland in 2020. It involved a breach of a
psychotherapy center whose sensitive data was leaked to the darknet and further
used in personalized extortion. Given the sensitivity of the personal data
involved, the breach was immensely traumatic for the persons affected; according
to media, it has been suspected that even suicides were committed due to the
breach. Therefore, it is no wonder that the breach has been a constant topic in
local news for the past four years. It has also gathered international media
attention. Early on during the crisis, when there was limited information
available, some speculated that the breach might have even been conducted by a
state actor in order to test the country's resilience. Later on, law enforcement
found the perpetrator to be a Finnish person with an already existing cyber
crime record from his teenage years. He was sentenced in 2024 to over six years
in prison due to the breach, extortion, and unauthorized release of private
information. The breach was the largest crime in Finland's history in terms of
the number of victims. The actual technical details that allowed the breach were
simple; the company's database was open to the Internet without a firewall and
it was protected by weak or even default passwords.

In the midst of the crisis, communication failed miserably; there were
widespread panic, speculations to all directions, and wild rumors. It was only
later on when public authorities, including law enforcement and the national
DPA, got involved that the crisis communication started to gain coherence and
reliability. There was no stealing of the thunder; it was only after the
perpetrator appeared in the darknet with the stolen data that the company was
forced to answer to inquiries from the media and public. Yet, there were plenty
of opportunities for self-disclosure. As the crisis unfolded, it became clear
that the company's management knew about the blatant cyber security
negligence. In fact, earlier breaches had occurred, but these had been brushed
under the carpet by the management. During the inquiries, the company's chief
executive officer (CEO) also contributed to the speculations and rumors by
stating that the breach might be an inside job. In addition, he and his business
associates sold their shares of the company in an attempt to salvage financial
losses. These attempts backfired. The company went bankrupt and the CEO was
later sentenced to a probation for a data protection crime. It was only after
the conviction that the CEO finally apologized for the victims about the serious
and foul incident. Regarding the GDPR, the company was fined a little over 600
thousand euros due to gross negligence of the regulation's various articles,
including Article~25 on the data protection by design and default, Article 32 on
the security of personal data processing, and the noted Articles 33 and 34 on
the obligations to inform a DPA and the persons affected. In terms of crisis
communication, it is further worth emphasizing that the company did not really
show remorse, accept responsibility, or offer solid advice on how the victims
should prepare for identity thefts and other consequences from the gross
violation of their privacy. Instead, the state was forced to organize crisis
support, including emotional and psychological help. Also volunteers contributed
to this helping effort, including during the law enforcement investigation and
later on when monetary compensations were pursued by the victims. All in all,
the Vastaamo breach is a showcase of almost all things that can go wrong in a
crisis management of a data~breach.

\subsection{The Helsinki City Breach}\label{subsec: helsinki}

In the early summer of 2024, it was reported that the city of Helsinki in
Finland had suffered a data breach. At the time of writing, the crisis is still
ongoing, but some lessons are still possible to draw from the breach. The breach
involved a compromise of the city's educational data. As is typical in data
breaches, the amount of persons affected increased step by step. According to
recent estimates, about 150,000 records were breached, including personal data
of past and present school pupils, their parents, and the city's
employees. According to media, the technical details that allowed the breach
showed gross negligence of cyber security; a vulnerability was unpatched for
almost two years in a server, although a patch was available from a vendor. By
exploiting the vulnerability, an attacker was able to breach a network drive
that contained over ten million documents. These documents contained personal
identifiers, including social security numbers, but also sensitive details of
the victims were present, including health records and details about special
educational needs of children. The unstructured storing of such a huge amount of
documents to a network drive further displays negligence of sound architectural
and infrastructural practices in information technology deployments.

The city's crisis communication strategy seemed to rely on positioning of the
city as a victim of a cyber crime. During an interview, the city's mayor made
this positioning explicit, but he neither apologized nor showed notable
remorse. Thus, it is no wonder that the public response has been highly
critical. According to a journalistic survey conducted by media, the parents
affected indeed exhibited deep anger toward the city, which, according to their
viewpoint, neither took responsibility nor provided clear advice on
mitigations. Also the city's crisis communication was widely criticized. No one
has apologized or offered empathy to the other, ``truer'' victims, the parents,
children, and employees affected by the data breach. Analogously to the Vastaamo
breach, the mitigation instructions were further largely outsourced to
third-parties, including the media in particular. On paper, everything went fine
with public authorities; law enforcement and the national DPA were notified as
required. As inconvenient facts started to appear in media, however, further
data protection negligence started to gain traction in public discourse. For
instance, the city was forced to admit that it had violated the GDPR's mandate
for fixed data retention times. Due to the network drive's configuration, the
city was also unable to determine whose personal data was specifically stolen;
hence, the city was unable to fulfill the GDPR's requirement of preferably
individualized notifications to the persons affected.  In addition, the likely
wrong choice of trying to play the victim card was supposedly made worse by the
fact that the city already had a bad reputation in information technology, given
an earlier crisis involving a severe dysfunction of the city's payment system
for its employees' salaries. Given this history and the public sector context,
it is understandable that the public discourse quickly further escalated into
politics. Among other things, political accountability was asked for and the
national DPA was forced to regret the Finland's odd decision to exclude public
sector organizations from the scope of the GDPR's administrative fines.

\subsection{The Kuopio City Accident}\label{subsec: kuopio}

In 2022 the city of Kuopio in Finland faced an accident in which personal data
of the city's employees was leaked through email. The data included salary
information, social security numbers, and information about belongings to labor
unions, among other things. The accident was simple; the city's human resource
department by mistake sent an Excel file containing the personal data to the
city's internal email list. After the incident, the city promptly notified the
national DPA, apologized for the employees, and, to some extent, offered also
sympathy to them. After fierce but constructive criticism from the employees
affected, the city further acknowledged that some of the data was likely
collected unnecessarily. The case is notable because it demonstrates that data
breaches can happen also by accident; not all breaches are about cyber crime.

\subsection{The Matriculation Examination Board Accident}\label{subsec: marticulation}

In 2018 it was reported in media that the Matriculation Examination Board, a
public sector organization that is responsible for granting high school
diplomas, accidentally put a system online that exposed personal data of about
six thousand young students. Once again, social security numbers were among the
data exposed. During an investigation conducted by the Board itself, the reason
behind the exposure was a subcontractor who had put a so-called staging server
online with the personal data exposed openly to the Internet. The case is worthy
of notice because it shows that poor software engineering practices can also
lead to data breaches; staging environments should not operate with pristine
personal data to begin with~\citep{Hjerppe19RE}. Fortunately, the investigation
also revealed that no one had accessed the data. While also notifying the media
through a self-disclosure, the Board handled the communication with the students
stylishly; each student was notified with a personal letter delivered through
conventional mail.

\subsection{The Prime Minister's Office Leak}\label{subsec: prime minister}

In 2015 sensitive email exchanges between a prime minister and several
high-profile civil servants were leaked from the the Prime Minister's
Office. The unencrypted email exchanges were about sanctions against Russia. The
perpetrator had offered the emails to media, and one magazine later on also
published these. During the incident response, the Office started an internal
examination and further notified the Finnish Security and Intelligence
Service. A former prime minister also commented the leak in media. During the
investigation, it became clear that the leak was not due to technical
exploitation. Otherwise, however, the investigation ended without a result. The
perpetrator was never caught. Although details were never published in media or
released as official statements, it seems likely that the leak was an inside job
within the Office. The motives behind the leak were likely political. Therefore,
this small-scale leak is worth noting because it highlights that data breaches
can occur also due to lapses in organizational security. Furthermore, the leak
has not been the only one affecting the Finnish central government in recent
years. For instance, in 2019 classified documents were leaked to media from the
Ministry of Foreign Affairs of Finland. Also this leak was related to a
politically hot topic, the Finnish family members of jihadi fighters in the
Al-Hawl refugee camp.

\subsection{The Traficom Breach}\label{subsec: traficom}

In mid-2024 another public sector breach was reported in Finland. Although the
details are not yet clear, the case is noteworthy because it involved a
supply-chain cyber attack. Namely, personal data from the Finnish Transport and
Communications Agency's automobile register was leaked through a cyber attack to
one of the register's third-party data processors. The attacker was able to
download about sixty thousand records through the attack. These included not
only car details but also personal identifiers, such as home addresses and
social security numbers of the car owners. The media has not reported how the
breach was noticed, including whether the processor notified the Agency, as
required by the GDPR. In terms of crisis communication, there has been a notable
information vacuum in addition to a lack of apologies from any party
involved. For instance, the Agency and other public authorities involved only
encouraged the data processor to voluntarily come into the public sphere to
answer to inquiries. Another point again relates to the mitigation instructions,
which have not been delivered in the crisis communication. During an interview,
a law enforcement officer investigating the breach correctly noted that data
breaches unfortunately happen and therefore it is important to prepare for the
consequences, but he offered no information sources on how the preparations
should be specifically done. This lack of information is regrettable because
many of the mitigations require adjusting privacy settings in information
systems of public sector organizations.

\subsection{The Forenom Breach}\label{subsec: forenom}

In 2020 a company that sells accommodation services was breached. The breach
affected various types of personal data, such as contact details and credit
ratings. The company communicated early through a self-disclosure, promptly took
responsibility, duly notified the national DPA, and apologized for the natural
persons affected by the breach. Despite this ``by the book'' communication
strategy, the company ended to an investigation by the DPA due to complaints
from several persons. During the investigation, it became clear that the breach
was made possible by a conventional structured query language (SQL) injection in
the company's software stack. Although no fines were imposed, the DPA further
found that the company had violated the GDPR's data minimization principle in
Article 5, the requirement for fixed data retention specified in the same
article, and the requirement for appropriate technical measures to ensure
information security, as specified in Article 32. The case is noteworthy because
it demonstrates that communication alone does not prevent public authorities
from taking action.

\subsection{The STT Attack}\label{subsec: stt}

In 2022 the Finnish News Agency (STT) got hit by a ransomware attack. Initially,
also a data breach was suspected because the criminal group behind the attack
claimed it had stolen all of the Agency's data. In fact, a data breach might
have been even worse than a ransomware attack because of the sensitivity of the
data involved, including data and data sources protected by journalistic
confidentiality. Fortunately, the claim turned out to be false. The Agency also
managed to stop the ransomware attack before its completion. Nor did the attack
stop the round-the-clock news production. Furthermore, the Agency stole the
thunder; a press release was immediately made already a day after the attack was
detected. Thereafter, several further updates were released. Also incident
response worked well; the Agency notified the national DPA and CSIRT, and
further made a criminal report to a local law enforcement unit. As a testimony
of good crisis management, the Agency later won a national information security
award from its openness and effective stakeholder collaboration during the
ransomware crisis.

\section{Discussion}\label{sec: discussion}

\subsection{Conclusion and Implications}

The eight cases presented exhibit an interesting mixture of different data
breach crisis communication strategies. Although existing research has been
critical about the applicability of the SCCT in the data breach
context~\citep{Knight20}, the theory's basic premise holds well; there is no
single universal communication strategy in data breach crises situations. As for
the theory's situational characteristics, particularly the question of how much
an organization is responsible for a data breach is relevant. It may cause
different interpretation problems during crisis management, especially given the
urgency and short response times in data breach crises. Robustly determining the
degree of responsibility is also necessary to comply with European
regulations. Most of the cases presented satisfied the incident notification
requirement toward public authorities. While all of the cases involved real
incidents, it is also possible that organizations nowadays notify public
authorities even in case an organization's responsibility is not entirely clear
at the time of the notifications. The strict deadlines in the GDPR and the NIS2
directive have likely contributed to such proactive notification
practices~\citep[cf.][]{Kasl21}. In any case, the law-imposed notifications can be
identified as a critical factor for a successful data breach crisis
communication strategy. These have been largely also ignored in existing crisis
communication research. Thus, further research is needed on these and the impact
of the regulations in general to crisis communication.

\begin{figure}[th!b]
\centering
\includegraphics[width=8cm, height=10cm]{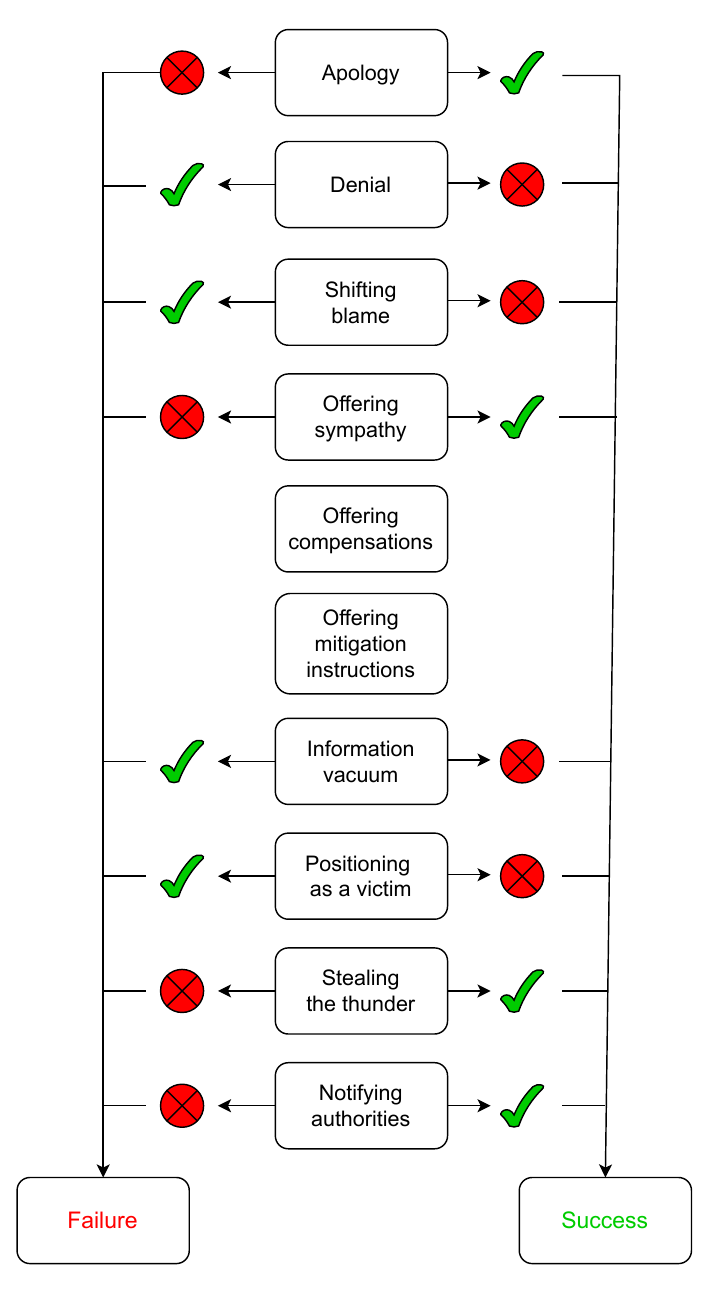}
\caption{Qualitative Process Tracing Results}
\label{fig: process results}
\end{figure}

As was discussed in Section~\ref{sec: crisis communication}, the regulatory
mandates affect many of the core theoretical premises in the literature. Due to
the short response times imposed, the stealing the thunder idiom is among these
premises. Most of the successful cases followed this idiom by not only notifying
public authorities but also informing the public and dealing with media
inquiries. As can be seen from the summary in Fig.~\ref{fig: process results},
the conventional wisdom from the literature was also otherwise followed in the
successful cases. While improvements are always possible, the organizations that
indicated successful crisis communication generally took responsibility,
apologized, and sometimes offered sympathy to the natural persons affected. The
unsuccessful organizations showed varying degrees of the reverse. Many of the
crisis communication missteps and mistakes were also identified as such in
media. This observation supports the argument raised in Section~\ref{sec:
  materials and methods}. For instance, the Helsinki city breach prompted also
a discussion in media about the city's more or less unsuccessful crisis
communication. Among other things, the city's attempt to position itself as a
victim was identified as a mistake, as could be expected also from the
literature. It likely further angered the other victims, the people
affected. The Helsinki city breach is also a good example otherwise because it
demonstrates that organizational impression management should apply prior to a
crisis.

The cases considered included also bad and ugly cases such as the Vastaamo
breach in which partial denial, downplaying, shifting blame, failing to comply
with regulations, and, in short, plain lying were all part of the crisis
communication ``strategy''. In line with the process tracing methodology, it
would be interesting to continue in further research by trying to identify the
particular mistake that was crucial in the sequence of failures. In other words,
there might have been a so-called tipping point after which it became impossible
to salvage the situation by the means of communication or otherwise. Again, the
failure to notify public authorities and responsibly cooperate with them would
be a good and perhaps the primary candidate for such a tipping point. On one
hand, therefore, it seems reasonable to state that failing at this task is a
sufficient but not a necessary condition for a failure in data breach crisis
communication. On the other, as the Forenom breach pinpoints, succeeding at this
task seems to be a necessary but not a sufficient condition for a success; even
a solid communication strategy may still cause regulatory action, although a
good crisis communication plan is likely to also reduce the probability of
administrative fines or even litigation.

As was pointed out in Section~\ref{subsec:
  notifying public authorities}, there is also a notable gap in the existing
crisis communication research in this regard. While there are some studies upon
which some practical advices can be drawn, the existing knowledge is very
limited regarding the collaboration with public authorities and its impact upon
crisis communication. It would be also possible to continue more general
theorizing along thee lines. For instance, drawing on \citeauthor{Lindblom59}'s
\citeyearpar{Lindblom59} classical concept of muddling through in organizations,
a concept of sliding has been used to describe incremental failure of industrial
innovation projects~\citep{Rehn12}. After identifying a tipping point, it might
be thus possible to theorize and empirically evaluate how and how fast a given
crisis communication process slid into a utter failure. It may also be that a
sliding process starts already from the very beginning. One lie might work, but
two lies may start a sliding process.

Also different information vacuums were present among the cases, suggesting that
some of the organizations involved perhaps did not have crisis communication
plans to begin with. Alternatively, there might have been internal issues within
the organizations, such as opting for a wait and monitor strategy, fearing legal
and regulatory implications, or a difficulty for the given crisis communicators
to convince an organization's management about a necessity to communicate
early~\citep[cf.][]{Claeys16}. The involvement of public authorities might offer
a further explanation. As was remarked, there may be situations, such as formal
investigations, that legally prevent an organization from communicating
thoroughly about a data breach crisis. This point too has been largely neglected
in existing research. The general societal role played by public authorities is
a little better known and theorized in the literature. In particular, the case
study results further indicate that public authorities have important auxiliary
functions beyond criminal and compliance investigations. In line with previous
results~\citep{Marynissen20}, in case of grievous data breaches, such as the
Vastaamo breach presented, public authorities and their communication also
contribute to lessening of panic and maintaining overall trust in a society. To
this end, also public authorities might consider evaluating or adjusting their
data breach crisis communication strategies. Although blaming public authorities
for slowness is a common political outcry~\citep{Palttala12}, timing of
communication is still one thing to consider. In many of the cases presented,
both the national DPA and law enforcement too were forced to enter the public
sphere as media kept pushing data breach narratives forward. Therefore, earlier
and a more proactive communication might be beneficial to the society at large.

An important further point is that none of the cases indicated offers of
voluntary monetary compensations or instructions for mitigative measures. Given
the descriptive tests for qualitative process tracing~\citep{Collier11,
  Mahoney12}, these observations thus reject or at least weaken hypotheses that
offering compensations or mitigations would be either sufficient or necessary
conditions for a successful data breach crisis communication. These and other
result might be explained by national characteristics. Although monetary
compensations were pursued through the justice system in the Vastaamo case, the
lack of voluntary compensation offers may generally relate to the Finnish
business culture. Although also contrary results have been
presented~\citep{LuomaAho17}, also the apparent difficulty or hesitance to offer
an apology in a data breach crisis situation may be due to a Finnish
organizational culture or, in the case of the public sector breaches, national
political culture. The same point applies with respect to offering sympathy or
empathy to the victims, which is mostly absent.

There is also a further important point to make about Finland and data
breaches. Namely, the leakage of social security numbers is often a serious
affair because these are still sometimes used incorrectly for authentication
instead of merely uniquely identifying persons, as is the purpose of the
numbers. The state recently changed the national data protection law to forbid
authentication with social security numbers, but the practice likely still
continues particularly in more informal settings such as phone calls to
organizations. Therefore, also the risks of financial fraud and identity thefts
are present in most of the cases considered. Even though the risks are mostly
about petty crime, such as ordering things from online stores or taking quick
loans with a stolen identity, these still cause a significant harm to the
persons affected already because resolving the issues take a long time and
requires activity from the persons themselves. In terms of communication,
therefore, it might be a good idea for the state or other public sector
organizations to improve their instructions for dealing with the
risks. Paradoxically, as already remarked, many of the relevant remedies involve
also adjusting settings in online systems of public sector organizations
themselves. More fundamentally, therefore, it might be a good idea to also
contemplate whether opt-in choices might be a better option than the opt-out
choices nowadays typically used by default in the public sector~context.

\subsection{Limitations}

Some limitations should be acknowledged. To begin with, a notable limitation is
that most of the theoretical premises and practical advices in Fig.~\ref{fig:
  process results} cannot be robustly identified as sufficient or necessary
conditions for a success or a failure in data breach crisis
communication. However, as was discussed in the previous section, this point
does not apply to all premises and advices. Some of these could be also
eliminated, suggesting a road in further research toward more parsimonious
theories and theoretical premises.

Then, generalizability should be acknowledged as a limitation. Although multiple
case studies were conducted, these cannot be considered as a representative
sample to describe data breach crisis communication generally in Finland, let
alone in Europe as a whole. However, the basic premise from the SCCT applies
also here; as no two crises are entirely similar, there is no one-size-fits-all
crisis communication strategy, and, therefore, it is important to rather learn
from good and bad examples~\citep{Horsley02}. Having said that, the truncated
sample resulting from the case selection based on polar types has its own
problems and potential biases~\citep{Collier96}. A~further limitation is that
the paper mostly shared the traditional organizational focus. This focus tends
to dismiss the interconnectedness in crises, omitting or downplaying important
social and institutional factors~\citep{Fredriksson14, Raupp19}. Although the
institutional factors were to some extent accounted for with the focus on
regulations and public authorities, highly severe data breach crises, such as
the Vastaamo breach, often require a whole-of-society response. Therefore,
further research should patch the limitation by considering a comprehensive
societal viewpoint to large-scale data breach crises. Such a viewpoint requires
also extending the scope of empirical material used. In this regard, a notable
further limitation is that the case studies were based on news articles in
traditional media, press releases from the organizations affected, and official
statements from public authorities. Nowadays, however, particularly the public's
response might be better gauged by data from social media. That said, neither
traditional media articles nor social media analytics can answer to a question
why particular crisis actors did what they did. Thus, further surveys and
interviews are required also in the data breach context. These should address
also victims of data breaches and inter-organizational communication with
stakeholders.

A further limitation is that most of the cases considered were traditional
personal data breaches. Thus, further insights are required on breaches
involving non-personal data, given that also the crisis communication strategies
may differ according to the type of data involved. On a related note, it should
be emphasized that while the GDPR does not apply to non-personal data, including
anonymized data, the NIS2 directive does not apply to organizations operating in
sectors defined as non-critical. This uneven treatment of
sectors~\citep{Ruohonen24I3E} may imply that some organizations fall through the
cracks; they may have no legal mandates to notify any public authority about a
breach. In a similar vein, there is no empirical knowledge on how organizations
have decided and acted upon the GDPR's different notification scenarios
discussed in Section~\ref{subsec: notifying public authorities}. Again, it may
be that some personal data breaches have fallen through the cracks due to
interpretative divergences or even potential errors in legal interpretation.

\subsection{New Openings for Further Work}

The conclusion reached and the limitations outlined furnish further research
possibilities. However, in what follows, the possibilities outlined are not
explicitly about the paper's conclusion or its limitations, both of which were
already sufficiently elaborated. Thus, in general, stakeholder coordination and
communication are among these possibilities; further work is required to better
understand how crisis communication occurs and how well it works when there are
several intersecting stakeholders involved, including public authorities who
should be notified beforehand. Do they bear their responsibilities or perhaps
blame each other? The work should also address crisis communication plans. Is a
plan negotiated jointly between an organization and its stakeholders or do the
latter operate on their own terms? Who bears the overall responsibility for
communication? Rather similar points apply to internal communication within
organizations. As was noted, it may well be that internal organizational
problems were behind the crisis communication difficulties in some of the cases
presented. Such problems could be perhaps addressed in further research by
asking organizations for their crisis communication plans. This approach would
allow to also compare the intended crisis communication with the actual
communication. Alternatively, it may be that organizations generally lack crisis
communication plans; such a result would be noteworthy on its own right.

Another new and novel organizational research topic is closely related to the
mentioned points about data breach crisis communication. According to the SCCT,
in most cases organizations should readily accept responsibility when a
situation requires to do so. However, only little is known about the practical
aspects behind taking responsibility in the data breach context. Who
specifically should take the responsibility? Should it be a CEO or perhaps an
organization's information technology department and its staff? The questions
are nontrivial because in some cases these extend beyond organizations' internal
management. As the Helsinki city data breach demonstrates, in some cases
responsibility may align with political accountability of elected or selected
representatives. Also legal issues may be involved, including criminal
liability, as the Vastaamo case again eagerly testifies.

The noted public sector mitigations would also offer a promising path for
further research. Although plenty of research has been done to examine people's
security awareness after data breaches, national and more specific surveys have
been lacking. Particularly the various opt-out options present in various
Finnish public sector organizations would need a closer examination. For
instance, it may well be that the general public is for the most part unaware of
the options, which is unfortunate already because the purpose of these is to
protect citizens from various harms, whether in terms of privacy or financial
matters. This presumption is reinforced by the case studies presented, none of
which indicated offers of reasonable mitigation instructions to the victims. As
awareness is closely related to communication, the topic does not diverge much
from the issues discussed in this paper Furthermore, more comparative research
is needed to determine whether the available mitigations are similar across
Europe. It may be that in some countries private sector options are more
prevalent than public sector ones. More research is also needed to evaluate how
effective the mitigations are in practice. Data from law enforcement on identity
thefts and related crimes would offer an ideal setup to examine this question.

Further research is also required to better understand the coping of persons
affected by data breaches. Although there is some existing
work~\citep{Gibson23}, only little is generally known about how the victims of
data breaches cope emotionally, psychologically, and otherwise. The Vastaamo
case is again a good albeit dismal example. This research theme aligns with the
noted whole-of-society crisis management approach because in many countries,
including Finland, public sector organizations offer different emergency
solutions for victims and other alarmed people, including chats and help through
phone emergency services. In addition to the coping itself, it would be
interesting to know how well the helping services have been communicated to the
public.

Finally, also more technical research is needed. Although the GDPR in particular
has stolen much of the attention in both research and practice, as was already
noted, data breaches are present also in standards such as the ISO/IEC 27000
information security standard. Further evaluation work is required to determine
how well the standard-imposed practices have been adopted and implemented in
organizations. Moreover, further research is required to better understand the
technical causes behind data breaches. Although technical details are often only
partially disclosed, even limited information on the probable causes can be
valuable for building preventive technical solutions and documenting best (and
worst) information technology practices, as has been also demonstrated
  in existing research~\citep{Saleem20}. Technical factors can be important
also for organizational crisis communication. As the Helsinki city case
demonstrates, technical factors and limitations can put crisis communicators
into a difficult position or even prevent an effective response to a data
breach.

\bibliographystyle{apalike}

\end{document}